\documentclass{aa} % for a referee version
\usepackage{graphicx}
\usepackage{txfonts}
\begin{document}
   \title{Solar-like oscillations in the metal-poor subgiant $\nu$~Indi}
   \subtitle{II. Acoustic spectrum and mode lifetime}

   \author{F. Carrier
          \inst{1,2}
          \and
          H. Kjeldsen
	  \inst{3}
	  \and
	  T.R. Bedding
	  \inst{4}
	  \and
	  B.J. Brewer	  
	  \inst{4}
	  \and
	  R.P. Butler
	  \inst{5}
	  \and
	  P. Eggenberger
	  \inst{2,6}
	  \and
	  F. Grundahl
	  \inst{3}
	  \and
	  C. McCarthy
	  \inst{5}
	  \and
%% 	  T.B. Nielsen
%% 	  \inst{3}
%% 	  \and
  	  A. Retter
  	  \inst{4,7}
  	  \and
  	  C.G. Tinney	  	  	  	  	  
	  \inst{8}
          }

   \offprints{F. Carrier}

   \institute{Instituut voor Sterrenkunde, Katholieke Universiteit Leuven, Celestijnenlaan 200D, 
	     B-3001 Leuven, Belgium\\
	     \email{fabien@ster.kuleuven.be}
	  \and
	     Observatoire de Gen\`eve, Universit\'e de Gen\`eve, 51 chemin des Maillettes,
             CH-1290 Sauverny, Switzerland
	  \and
	      Department of Physics and Astronomy, University of Aarhus,
             DK-8000 Aarhus C, Denmark\\
             \email{hans@phys.au.dk,fgj@phys.au.dk,tbn@phys.au.dk}
	  \and
	      School of Physics A28, University of Sydney, NSW 2006,
             Australia\\
	     \email{bedding@physics.usyd.edu.au,brewer@physics.usyd.edu.au}
	  \and
%	      Laboratoire d'Astrophysique de Marseille, Traverse du Siphon, BP 8,
%	     13376 Marseille Cedex 12, France
%	  \and
             Carnegie Institution of Washington, Department of Terrestrial Magnetism, 
	     5241 Broad Branch Road NW, Washington, DC 20015-1305\\ 
	     \email{paul@dtm.ciw.edu, chris@dtm.ciw.edu}
	  \and   
	     Institut d'Astrophysique et de G\'eophysique, Universit\'e de Li\`ege, 
	     17 All\'ee du 6 Ao\^ut, B\^at. B5c, B-4000 Li\`ege, Belgium\\
	     \email{eggenberger@astro.ulg.ac.be}
          \and
	     Department of Astronomy and Astrophysics, PennSylvania State University, 525 Davey Lab, University Park,
	     PA 16802-6305\\
	     \email{retter@astro.psu.edu}
	  \and
             Anglo-Australian Observatory, P.O.\,Box 296, Epping, NSW
             1710, Australia\\
	     \email{cgt@aaoepp.aao.gov.au}
          }
   \date{Received ; accepted }

% \abstract{}{}{}{}{} 
% 5 {} token are mandatory
 
  \abstract
   {Convection in stars excites resonant acoustic waves which depend on the sound speed
inside the star, which in turn depends on properties of the stellar interior. Therefore, asteroseismology is 
an unrivaled method to probe the
internal structure of a star.}
  {We made a seismic study of the metal-poor subgiant star \object{$\nu$~Indi} with the goal of constraining its interior structure.}
   {Our study is based on a time series of 1201 radial velocity measurements spread over 14 nights obtained from two sites,
   Siding Spring Observatory in Australia and ESO La Silla Observatory in Chile.}
   {The power spectrum of the
high precision velocity time series clearly presents several identifiable peaks between 200 and
500\,$\mu$Hz showing regularity with a large and small spacing of $\Delta\nu$\,=\,25.14\,$\pm$\,0.09\,$\mu$Hz and 
$\delta\nu_{02}$\,=\,2.96\,$\pm$\,0.22\,$\mu$Hz at 330\,$\mu$Hz. 
Thirteen individual modes have been
identified with amplitudes in the range 53 to 173\,cm\,s$^{-1}$. The mode damping time is estimated to be about 16~days (1-$\sigma$ range between 9 and 50 days),
substantially longer than in other stars like the Sun, 
the \object{$\alpha$~Cen} system or the giant \object{$\xi$~Hya.}}
   {}

   \keywords{Stars: individual: $\nu$~Indi --
          Stars: oscillations -- Stars: variables: general -- Stars: interiors
               }

   \maketitle
%
%________________________________________________________________

\section{Introduction}

The analysis of the oscillation spectrum provides an unrivaled method to probe the stellar
internal structure.  The frequencies of these oscillations depend on the sound speed inside the
star, which in turn depends on density, temperature, gas motion and other properties of the stellar
interior. High-precision spectrographs have led in recent years to a rapidly growing list of 
solar-like oscillation detections. 

The recent achievements are reviewed by Bedding \& Kjeldsen (\cite{bk06}, \cite{bk07}).
%% except for the new detection of p-modes on the K0 dwarf \object{70~Oph~A}
%% (Carrier \& Eggenberger \cite{ce06}).
All of the previous detections of oscillations were made on stars with
metallicities close to solar.
We recently reported observations of the metal-poor subgiant star \object{$\nu$~Indi} made in both 
Chile with the spectrograph CORALIE
and in Australia with UCLES (Bedding et al. \cite{bbce06}, hereafter Paper I). These observations covered a 14 nights
time span and had the benefit of two-site coverage.

In Paper I, we described the radial velocity time series and calculated the corresponding power spectrum weighted by the 
measurement uncertainties. We reached a noise floor of 14.9\,cm\,s$^{-1}$ in the combined amplitude spectrum, which is quite high compared to other
studies with the same instruments (see e.g. Bouchy \& Carrier \cite{bc02} and Bedding et al. \cite{bkbe04}). This difference can be explained by the extremely low metallicity 
(about 3\% of solar) and the faint magnitude (m$_V$\,=\,5.28) of \object{$\nu$~Indi}. Oscillation modes
are present around 0.32\,mHz with a maximal amplitude per mode of 95\,cm\,s$^{-1}$. A large frequency separation was determined
by autocorrelation and Bayesian methods to be 24.25\,$\pm$\,0.25\,$\mu$Hz.
   \begin{figure*}[!Ht]
   \resizebox{\hsize}{!}{\includegraphics{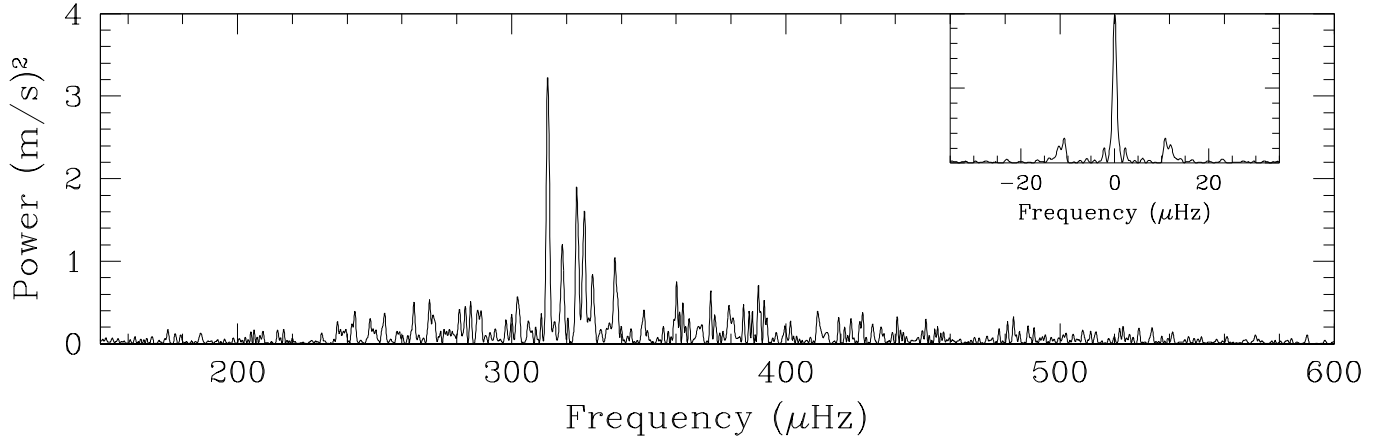}}
   \caption{Power spectrum of the radial velocity measurements of \object{$\nu$~Indi}. The inset shows the spectral window.}
              \label{figtf}%
    \end{figure*}

In this paper, we will extend the analysis by extracting individual oscillation frequencies, by measuring the large and 
small separations, and by estimating the mode lifetimes.

%__________________________________________________________________

\section{Frequency analysis}
\label{fa}
The power spectrum of the time series, shown in Fig.~\ref{figtf}, exhibits a series of peaks between 200 and 550\,$\mu$Hz.
In solar-like stars, p-mode oscillations are expected to produce a
characteristic comb-like structure in the power spectrum with mode
frequencies
$\nu_{n,\ell}$ reasonably well approximated by the asymptotic
relation (Tassoul \cite{tassoul80}):
\begin{eqnarray}
\label{eq1}
\nu_{n,\ell} & \approx &
\Delta\nu(n+\frac{\ell}{2}+\epsilon)-\ell(\ell+1) D_{0}\;.
\end{eqnarray}
Here $D_0$ (which equals $\frac{1}{6} \delta\nu_{02}$ if the 
asymptotic relation holds exactly) and $\epsilon$ are sensitive to the sound speed near the core and in
the surface layers, respectively. 
The quantum numbers $n$ and $\ell$ correspond to the radial
order and the angular degree of the modes, and $\Delta\nu$ and
$\delta\nu_{02}$
are the large and small spacings.
To fit to this relation,
the large separation was measured in Paper I by autocorrelation and Bayesian methods and was found to be 24.25\,$\pm$\,0.25\,$\mu$Hz.

Note that a subgiant such as \object{$\nu$~Indi} is expected to show
substantial deviations from the regular comb-like structure described
above.  This is because some mode frequencies, except for 
$\ell=0$, could be shifted from their usual regular spacing by avoided crossings
with gravity modes in the stellar core (also called `mode bumping') (see e.g. Christensen-Dalsgaard et al. \cite{cd95} and Fernandes \& Monteiro \cite{fm03}).  We
must keep in mind the possibility of these mixed modes when
attempting to identify oscillation modes in the power spectrum.

The next step is to measure the frequencies of the strongest peaks in the
power spectrum. 
%
%% The stochastic nature of solar-like oscillations implies that a timestring
%% of radial velocities cannot be expected to be a set of coherent
%% oscillations and can therefore not be reproduced perfectly by a sum of
%% sinusoidal terms unless we include a large .
%% The mode lifetimes will be determined in Section~\ref{lifescatter}. 
%
% Comment from Tim and Hans: Any signal can be reproduced by a sum of sine
% waves, but we just need a lot more of them if the signal is stochastic.
% In this case, where the modes are not very resolved, we *can* match the
% time series with a small number of terms.
%
The frequencies were  extracted using an iterative algorithm 
which identifies the highest peak 
between 150 and 600\,$\mu$Hz and subtracts it from the time series (see e.g. Carrier et al. \cite{cb03} and Kjeldsen et al. \cite{kbbe05}).
In the present analysis we are only marginally affected by the window sidebands compared to mono-site observations analysis. The highest amplitude modes 
are therefore expected to be located  at their correct position and they do not need to be shifted by the daily aliases (Kjeldsen et al. \cite{kbbe03}, Carrier et al.
\cite{ceb05,cedw05}). In total, 27 frequencies were extracted having an amplitude 3 times greater than the noise. These amplitudes are
measured from the best-fitting sinusoid, whereas the noise is computed as the mean in the amplitude spectrum, which is the square root of the power spectrum, at high frequencies.
Note that the noise 
level at high frequencies (600--1000\,$\mu$Hz) has a value of 14.9\,cm\,s$^{-1}$ (Paper I) and is assumed constant on the whole range 150--600\,$\mu$Hz.
   \begin{figure}
   \resizebox{\hsize}{!}{\includegraphics{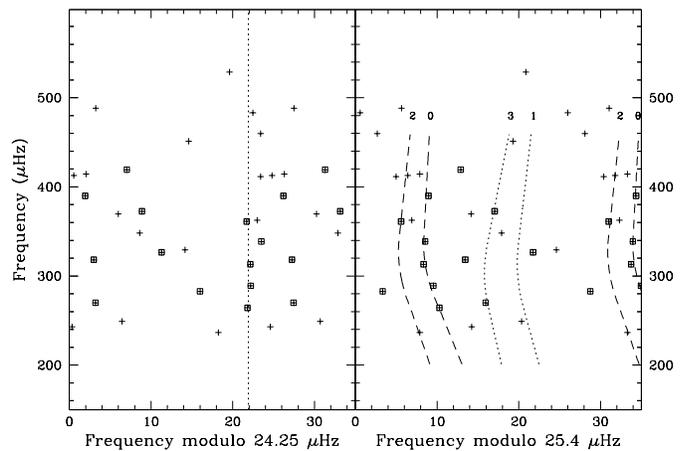}}
   \caption{Echelle diagrams for two different values of the large separation. The modes with a SNR greater than 4 are indicated with a square, and the ones
   with a SNR greater than 5 are indicated in boldface. {\bf Left:} Echelle diagram where 24.25\,$\mu$Hz is used as a value for the large separation (see Paper~I).
   {\bf Right:} Echelle diagram with a large separation of 25.4\,$\mu$Hz. The lines are just meant as a "guide".}
              \label{figdech1}%
    \end{figure}

All frequencies with a signal-to-noise ratio (SNR) above 3 were then placed in an echelle diagram, without shifting any of them, constructed 
with the value for the large
separation found in Paper I (see Fig.~\ref{figdech1} left). 
This diagram does not seem to show a classical
asymptotic structure, since only one vertical ridge can be seen (dotted
line at 22\,$\mu$Hz).
A possible way to explain such a graphic is to associate the ridge at 22\,$\mu$Hz to $\ell\,=\,2$ modes
and the quasi-line near 27\,$\mu$Hz to $\ell\,=\,0$ modes. This identification is not very convincing and
would imply a small separation of 5\,$\mu$Hz, which is too large for this kind of star.

Another way to understand this diagram is to increase slightly the large
spacing and allow curvature along the frequency axis (see
Fig.~\ref{figdech1} right).  Indeed, large separation determination methods
based on the asymptotic relation (as the autocorrelation presented in
Paper~I), do not take into account any curvature. 
With curvature, there is no correct value for the large spacing, it is a function
of frequency.
With a large separation of 25.4\,$\mu$Hz, as in the right
panel, we can identify ridges corresponding to $\ell\,=\,0$ and
$\ell\,=\,2$.  These are marked with dashed lines.

With the ridges for $\ell = 0$ and $2$ marked as shown, we then used the
asymptotic relation to calculate the expected positions for $\ell = 1$ and
$3$.  These are shown by dotted lines in the right panel of
Fig.~\ref{figdech1}.  
We can see that two strong peaks lie exactly on the line for $\ell=3$,
giving us some confidence that we have found the correct solution.  We also
see an absence of peaks on the $\ell\,=\,1$ line and a few strong peaks
that do not lie on any of the lines.  This is consistent with the presence
of mixed $\ell\,=\,1$ modes that are shifted by avoided crossings, as
described above.

\begin{figure}
   \resizebox{\hsize}{!}{\includegraphics{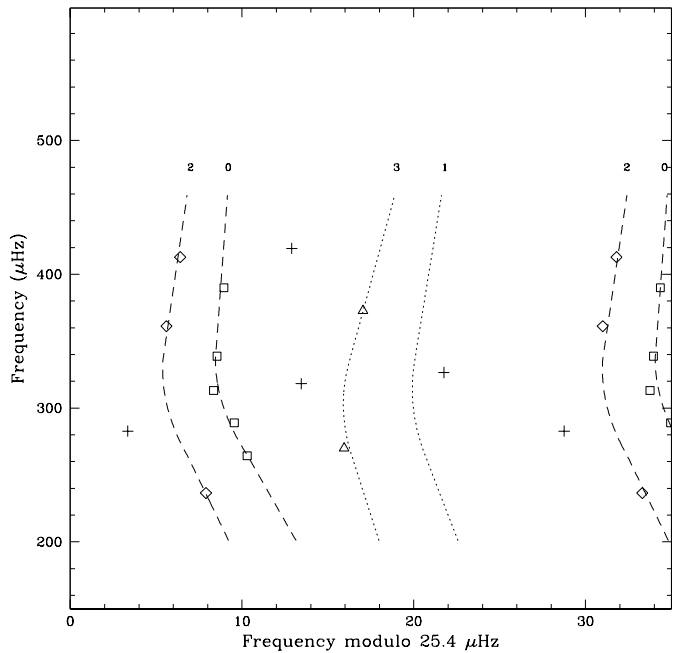}}
   \caption{Echelle diagram for the frequencies given in Table~\ref{modes.sn} and~\ref{modes}. The lines are just meant as a "guide".
   Squares, crosses, rhombuses and triangles represent the $\ell$\,=0, 1, 2 and 3 modes respectively.}
              \label{dech}%
    \end{figure}

Based on Fig.~\ref{figdech1}, several modes were identified in
\object{$\nu$~Indi}. The frequency, amplitude, SNR of these modes are given
in Table~\ref{modes.sn} with also the identified degree. Since the
amplitude of the modes in an iterative fitting algorithm depends slightly
on the selection order, a least squares fit was applied at the end to
determine all amplitudes simultaneously.  We retained all peaks with a SNR
greater than 4.  We also added two weaker peaks that were in good agreement
with the $\ell\,=\,2$ ridge (see Table~\ref{modes.sn}).  It is expected in
evolved stars that $\ell\,=\,1$ modes are the first to be mixed modes and
do not follow the asymptotic relation.  Thus, all large peaks which can not
be determined as $\ell$\,=\,0, 2 or 3 are listed as $\ell\,=\,1$ modes.
$\ell\,=\,3$ modes are also less reliable. Finally, all the selected modes
are shown in the echelle diagram on Fig.~\ref{dech} and listed in
Table~\ref{modes}.

\begin{table}
\caption[]{Amplitude and frequency for individual oscillation modes and their identification.
%
%% The error on the frequencies and amplitudes are calculated according to
%% Montgomery \& O'Donoghue (\cite{mo99}), assuming an infinite mode lifetime.
%% [we suggest not to show these because they are not realistic]
The mean noise in the amplitude spectrum is 11.9\,cm\,s$^{-1}$.}
\begin{center}
%% \begin{tabular}{rcccc}
\begin{tabular}{rccc}
\hline
\hline
%% \multicolumn{1}{c}{Frequency} & Amplitude & SNR  & Mode ID & Error on Frequency\\
%% \multicolumn{1}{c}{$[\mu$Hz$]$} &   cm\,s$^{-1}$ &  & $\ell$ &  $\mu$Hz\\
\multicolumn{1}{c}{Frequency} & Amplitude & SNR  & Mode ID \\
\multicolumn{1}{c}{$[\mu$Hz$]$} &   cm\,s$^{-1}$ &  & $\ell$ \\
\hline
313.1 & 172  & 11.6 & 0  \\ % & 0.04 \\
372.6 & 108  &  7.2 & 3? \\ % & 0.06 \\
361.3 & 105  &  7.1 & 2  \\ % & 0.06 \\
318.2 & 103  &  6.9 & 1? \\ % & 0.06 \\
389.9 & 100  &  6.7 & 0  \\ % & 0.06 \\
326.5 &  97  &  6.5 & 1  \\ % & 0.06 \\
264.3 &  91  &  6.1 & 0  \\ % & 0.07 \\
338.8 &  89  &  6.0 & 0  \\ % & 0.07 \\
282.7 &  73  &  4.9 & 1? \\ % & 0.09 \\
288.9 &  69  &  4.6 & 0  \\ % & 0.09 \\
269.9 &  69  &  4.6 & 3? \\ % & 0.09 \\
236.5 &  54  &  3.6 & 2  \\ % & 0.11 \\
412.8 &  54  &  3.6 & 2  \\ % & 0.12 \\
\hline
\end{tabular}
\end{center}
\label{modes.sn}
\end{table}

\begin{figure}
   \resizebox{\hsize}{!}{\includegraphics{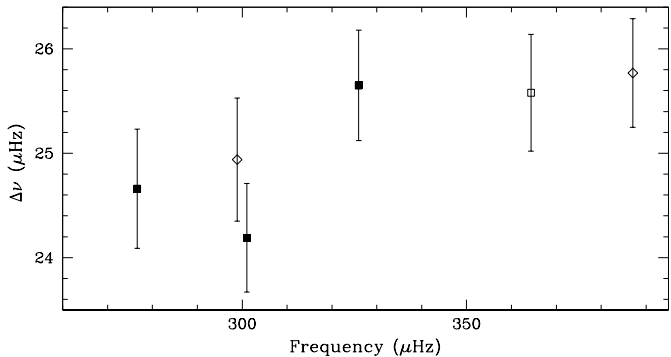}}
   \caption{Large spacing $\Delta\nu$ versus frequency for p-modes of degree $\ell$\,=\,0 ($\square$) and $\ell$\,=\,2 ($\diamond$). Open symbols correspond
   to large spacing averages taken between non-successive modes.}
              \label{large}%
    \end{figure}

\begin{table}
\caption[]{Frequency (in $\mu$Hz) and mode identification for individual oscillation modes.
The errors on the frequencies are given in brackets (10$^{-1}$ $\mu$Hz) according simulations described in Section~\ref{erfa}. The identification of frequencies in brackets 
is less secure.}
\begin{center}
\begin{tabular}{rcccc}
\hline
\hline
n & $\ell$\,=\,0 & $\ell$\,=\,1 & $\ell$\,=\,2 & $\ell$\,=\,3 \\
\hline
 7 &             &              & 236.5(5)   &              \\
 8 &             &              &              & ( 269.9(4) ) \\
 9 & 264.3(4)   &              &              &              \\
   &             &  ( 282.7(4) ) &              &              \\
10 & 288.9(4)   &              &              &              \\
   &             &  ( 318.2(4) ) &              &              \\
11 & 313.1(3)   &  326.5(4)   &              &              \\
12 & 338.8(4)   &              & 361.3(3)    & ( 372.6(4) )  \\
13 &             &              &              & 	      \\
14 & 389.9(4)   &	        & 412.8(4)   &              \\
\hline
$\Delta\nu$ & 25.14 & & 25.14 & \\
%%          & $\pm$ 0.14 & & $\pm$ 0.11 & \\
            & $\pm$ 0.22 & & $\pm$ 0.18 & \\
\\
%% \multicolumn{1}{l}{$\Delta\nu$} & \multicolumn{4}{c}{25.14 $\pm$ 0.09 $\mu$Hz at 0.33\,mHz}\\
\multicolumn{1}{l}{$\Delta\nu$} & \multicolumn{4}{c}{25.14 $\pm$ 0.14 $\mu$Hz at 0.33\,mHz}\\
\hline
\end{tabular}\\
\end{center}
\label{modes}
\end{table}
Based on Fig.~\ref{dech} and on the frequencies given in Table~\ref{modes}, the large  and small separation are deduced taking into
account only the $\ell$\,=\,0 and 2 modes, as $\ell$\,=\,1 modes do not follow the asymptotic relation and $\ell$\,=\,3 ones
do not have a secure identification. The large separation has a value of 25.14\,$\pm$\,0.14\,$\mu$Hz at 0.33\,mHz which corresponds
to the weighted average of individual values given for each degree (see Table~\ref{modes}). The variation of the large separation
versus frequency is presented in Fig.~\ref{large}.

For the small separation, there are no pairs of adjacent $\ell$\,=\,0,~2
modes, and so missing $\ell$\,=\,0 modes were linearly interpolated.  
%
%% In total, three value for the distance between $\ell$\,=\,0 and
%% $\ell$\,=\,2 can be calculated: 3.10, 3.09 and 2.71\,$\mu$Hz, resulting in
%
We find a value of $\delta\nu_{02}$\,=\,3.0\,$\pm$\,0.3\,$\mu$Hz at
0.33\,mHz.  It is not possible to calculate a reliable value for 
$\delta\nu_{13}$ since $\ell$\,=\,1 modes do not at
all follow the asymptotic relation, and since the identification of
$\ell$\,=\,3 modes could be in error. In fact, one could also identify the
$\ell$\,=\,3 as $\ell$\,=\,1 modes. This fact should be remembered when
comparing the observed and theoretical frequencies.

\section{Mode lifetimes}

We have used two methods to estimate the mode lifetimes from our data.  The
first uses the distribution of mode amplitudes and the second uses the
scatter of the frequencies about the ridges in the echelle diagram.
In both cases, we calibrated the results using simulations.

\subsection{Mode lifetimes by amplitude comparison}

\begin{figure}
   \resizebox{\hsize}{!}{\includegraphics{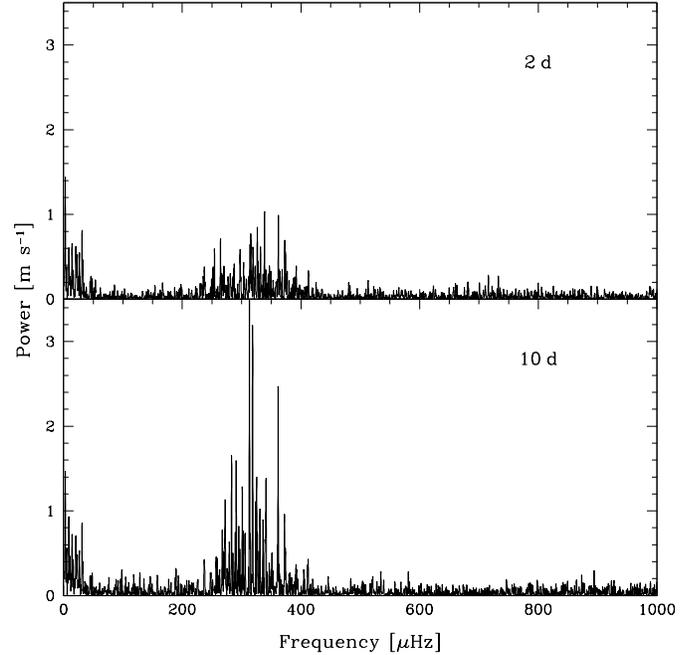}}
   \caption{Examples of two simulations of the \object{$\nu$~Indi} time
   series using mode lifetimes of 2~days (top) and 10~day (bottom). Higher
   peaks are in general found for the longer lifetime.}
              \label{simu1}%
    \end{figure}

The idea is to compare the amplitudes of peaks in the power spectrum with
simulations that have the same amplitude envelope as the observations and a
range of mode lifetimes.  The simulator for generating solar-like
oscillation time series is similar to that described by De Ridder et al.\
(\cite{drak06}).  For the input amplitudes, we used the smoothed amplitude
spectrum show in Fig.~7 of Paper~I.  We ran ten simulations for each of six
mode lifetimes (1, 2, 5, 10, 20 and 50 days).  Two examples, with mode
lifetimes of 2\,d and 10\,d, are shown in Fig.~\ref{simu1}.  We see a
significant difference between them, which can be used to estimate the mode
lifetime.

In order to do this, we extracted the 7 highest peaks from each simulated
power spectrum (using iterative fitting) and compared their amplitudes with
those extracted from the observed series.  As a simple measure of the
degree of agreement, we adopted the following method.  For a given
simulation, we compared the highest peak with the highest peak in the
observation.  We then compared the second highest simulated peak with the
second highest observed peak, and so on.  The fraction of peaks in the
simulation that were above the observed ones, in this pairwise comparison,
gave the measure of agreement, with 50\% indicating the best agreement.

\begin{figure}
   \resizebox{\hsize}{!}{\includegraphics{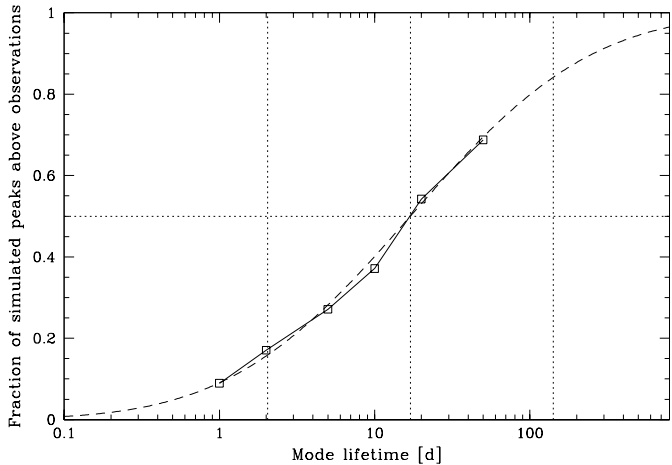}}
   \caption{Calibration of the
simulations to determine mode lifetime from the peak amplitudes.  It shows
the fraction of modes in the simulations that have amplitudes above the
observed mode amplitudes. The dashed curve is a normally distributed
function that presents the best fit.  Vertical dashed lines show the
resulting mode lifetime and the values at $\pm 1\sigma$.}
              \label{simu2}%
    \end{figure}

The results are shown in Fig.~\ref{simu2}, which shows the fraction of
simulated peaks that were above the observations.  We can read off the most
likely value of the mode lifetime, corresponding to a fraction of 0.5, to
be 17~days.  However, this measurement is not very precise.  The dashed
curve is the cumulative distribution of a normal distribution, and we see
that the one-$\sigma$ range for the mode lifetime is 2 to 141\,days, or in
the logarithmic units:
\begin{equation}
\log _{10} ( \tau / d) = 1.23 \pm 0.92.
\end{equation}

\subsection{Mode lifetime by frequency scatter}
\label{lifescatter}

The other way to extract information on mode lifetimes is to examine the
scatter of measured frequencies about the ridges in the echelle diagram.
For these observations, this is only possible for the five $\ell = 0$
modes.  If the modes were coherent (i.e. pure sinusoids with infinite
lifetimes), we would be able to measure them with a typical accuracy of
0.1\,$\mu$Hz.  The observed scatter is much greater than this, which we
interpret, at least for $\ell$\,=\,0 modes, as reflecting the finite
lifetimes of the oscillation modes.  

As is well established for the Sun, the power spectrum of a stochastically
excited oscillation that is observed for long enough will display a series
of closely spaced peaks under a Lorentzian profile, the width of which
indicates the mode lifetime (e.g. Toutain \& Fr\"ohlich \cite{tf92}).  If
the observations are not long enough to resolve the Lorentzian profile,
then the effect of the finite lifetime is to randomly shift each
oscillation peak from its true position by a small amount. This is
responsible for the scatter of the frequencies.

   \begin{figure}
   \resizebox{\hsize}{!}{\includegraphics{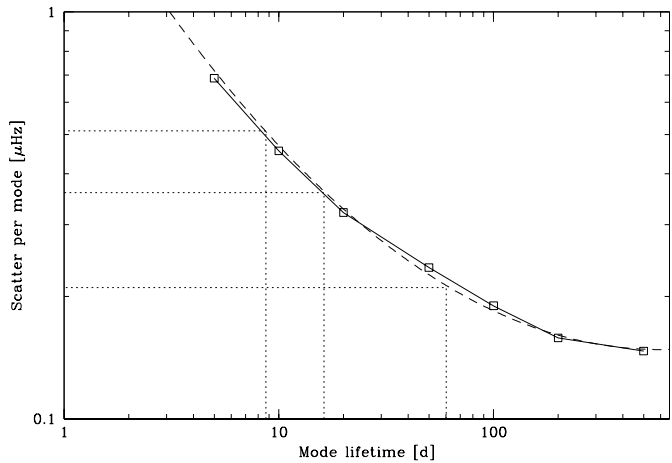}}
   \caption{Relation between scatter per mode and mode lifetime in
   days. The curve is a fit to points that came out of 700 individual
   simulated time series data. The mode lifetime corresponding to the best
   fitted frequency scatter is shown together with values for $\pm
   1\sigma$.}
              \label{simu3}%
    \end{figure}
    
The scatters are determined by assuming that peaks are unaffected by
the rotation of the star 
and that the large spacing varies only slightly between three
consecutive detected modes.  For each measured $\ell$\,=0 frequency, except
those at the ends of the ridges, we calculated the difference between the
measured frequency and that expected from linear interpolating the
positions of the two nearest neighbors.  This quantity was then transformed
to the rms scatter of that peak about its expected position, in order to
easily compare it with simulations (see Kjeldsen et al. \cite{kbbe05}).  In
this way, three values of scatter per mode could be calculated,
%% (see Table~\ref{tsimu1}) 
which have a mean of 0.36\,$\mu$Hz $\pm$ 0.15\,$\mu$Hz, for an average
observed amplitude of 95\,cm\,s$^{-1}$.  

In order to calibrate the relation between frequency scatter and mode
lifetime, we ran 700 simulations using the same sampling and statistical
weighting as the real time series. The results are shown in
Fig.~\ref{simu3} and the best fit corresponds to a mode lifetime of 16.2
days (between 8.7 and 60.8 days at 1\,$\sigma$). Expressed in logarithmic
units, we get:
\begin{equation}
\log _{10}  ( \tau / d) = 1.21^{+0.57}_{-0.27}
\end{equation}
Combining this measurement of mode lifetime with the previous one, and
assuming that they are independent estimates, we finally get:
\begin{equation}
\log _{10}  ( \tau / d) = 1.21^{+0.49}_{-0.26}
\end{equation}
or 16\,d (1-$\sigma$ range between 9 and 50 days). Note that the distribution about the mean is very asymmetric
which implies that the upper limit is quite tentative. 
Moreover, the quoted lifetime is very close to the bin width (resolution)
in the power spectrum and the mean can thus be influenced by the observation time length. However, the lower limit of 9\, days is quite secure and valuable.

{An interesting work to test the accuracy of the lifetime 
would be to compare the lifetime deduced from the scatter of the $\ell$\,=\,0 and  $\ell$\,=\,2 modes. 
Unlike $\ell$\,=\,0 modes, $\ell$\,=\,2 modes can be affected by the rotational splitting (v\,$\sin$\,i of 2.6\,km\,s$^{-1}$ determined from the CORALIE spectra)
and their scatter should be larger. However, the number and the non-consecutive position at all of such modes (see Fig.~\ref{dech}) do not allow us to do it.
%% \begin{table}
%% \caption[]{Amplitude, frequency and scatter due to the finite lifetime for individual $\ell$\,=\,0 modes.}
%% \begin{center}
%% \begin{tabular}{rcccc}
%% \hline
%% \hline
%% Amp & Frequency & Modes & Difference   & Scatter per mode\\
%% cm\,s$^{-1}$ & $\mu$Hz  & n & $\mu$Hz &  $\mu$Hz \\
%% \hline
%% 80.9 & 264.28   &   9          &         &      \\
%% 68.7 & 288.94   &   10         &  0.24   &  0.20   \\
%% 179.9& 313.13   &   11         &  -0.73  &  0.60     \\
%% 66.9 & 338.78   &   12         &  0.05   &  0.04   \\
%% 89.1 & 389.94   &   14         &         &  \\
%% \hline
%% \multicolumn{5}{l}{scatter per mode: 0.36\,$\mu$Hz $\pm$ 0.15\,$\mu$Hz} \\
%% \hline
%% \end{tabular}\\
%% \end{center}
%% \label{tsimu1}
%% \end{table}

\subsection{Frequency accuracy}
\label{erfa}

Now that we have determined the mode lifetime, we can estimate the
uncertainties in the mode frequencies.  To do this, we made 850 additional
simulations using several different amplitudes and assuming noise and
sampling similar to the \object{$\nu$~Indi} series in order to determine
the scatter of the detected modes.  The result of the simulations allowed
an estimate of the accuracy for the individual detected frequencies,
including the effect of the mode lifetime, and these are given in
Table~\ref{modes}.

\section{Conclusions}

Our observations of \object{$\nu$~Indi} from two sites have allowed us to
identify 13 oscillation modes.  The mode identification was complicated by
the apparent effects of avoided crossings.  We found a large and small
separation of $\Delta\nu$\,=\,25.14\,$\pm$\,0.09\,$\mu$Hz and
$\delta\nu_{02}$\,=\,2.96\,$\pm$\,0.22\,$\mu$Hz at 330\,$\mu$Hz.

Based on the scatter of the observed frequencies, we inferred a mode
lifetime of about 16\,days (1-$\sigma$ range between 9 and 50 days).
The lower limit of 9\,days is secure whereas the mean and upper limits values 
can be influenced by the observation time length and have to be taken with caution.
Note that the lifetime of \object{$\nu$~Indi}, even its lower limit of 9\,days, 
is far greater than previous detections
on other stars as the Sun (Chaplin et al. \cite{chaplin}), the dwarf stars
\object{$\alpha$~Cen~A} and B (Kjeldsen et al. \cite{kbbe05}) and the giant
\object{$\xi$~Hya} (Stello et al. \cite{stello}). This new measurement of
damping time, which is until now a poorly known parameter, is very
different from the previous ones and thus present a challenge to
theoretical models.  The theoretical study of $\nu$~Indi, with
asteroseismic and non-asteroseismic constraints, is postponed to a third
paper.

\begin{acknowledgements}
This work was supported financially by the Swiss National Science
Foundation, the Australian Research Council, the Danish Natural Science
Research Council, the Danish National Research Foundation through its
establishment of the Theoretical Astrophysics Center, and by a research
associate fellowship from Penn State University. We further acknowledge
support by NSF grant AST-9988087 (RPB) and by SUN Microsystems.
\end{acknowledgements}

\end{document}